\documentclass[11pt]{article}
\usepackage{psfig}
\begin{document}
\thispagestyle{empty}
\begin{center}
{\LARGE\bf On the Universality of Matrix Models}\\[0.3cm]
{\LARGE\bf for Random Surfaces}\\[1.0cm]
{\Large Antje Schneider}\\[0.2cm]
{\large FB 7, Theoretische Physik\\
Universit\"at Essen\\
D--45\,117 Essen}\\[0.5cm]
{\Large Thomas Filk}\\[0.2cm]
{\large Institut f\"ur Theoretische Physik\\
Universit\"at Freiburg \\
Hermann-Herder-Str.\ 3 \\
D--79\,104 Freiburg}
\end{center}
\vspace{1.5cm}

\begin{center} {\Large\bf Abstract}  \end{center}
\vspace{0.3cm}

We present an alternative procedure to eliminate irregular contributions 
in the perturbation expansion of $c\!\!=\!\!0$--matrix models
representing the sum over triangulations of random surfaces, thereby  
reproducing the results of Tutte \cite{Tutte} and Br\'{e}zin et
al.\,\cite{bipz} for the planar model. The advantage of this method is that  
the universality of the critical exponents can be proven from general  
features of the model alone
without explicit determination of the free energy
and therefore allows for several straightforward generalizations including
cases with non-vanishing central charge $c< 1$.        
\vspace*{3.0cm}

\mbox{~}
\hfill \begin{minipage}[t]{4cm}
\begin{center}
Universit\"at Essen \\
THEP ??\\
September 1998
\end{center}
\end{minipage}

\newpage
\setcounter{page}{1}

\section{Introduction}
The use of matrix models for the description and solution of theories
of 2-dimensional quantum gravity coupled to matter with conformal
weight $c\leq 1$ is well established (see e.g.\ the reviews
\cite{rev} and references therein). Critical exponents agree with those
found in the continuum description where methods of conformal quantum
field theory can be applied \cite{cont}.
The fact that this agreement is by
no means trivial is mostly overlooked, although the equivalence of the
theory of continuous 2-dimensional surfaces and the theory of abstract 
(combinatorical) triangulations with respect to their critical behaviour
is based on many assumptions and indeed is presumably wrong for
$c>1$. For the case of pure gravity ($c\!=\!0$) an integration over all 
metrics on a 2-dimensional surface modulo diffeomorphisms is replaced
by a summation over abstract triangulations T, 
which are defined merely by the adjacency properties of their
points:
\begin{equation}
Z^{\rm cont}_{\rm pure}=
   \int\frac{{\cal D}g_{\alpha\beta}}{\mbox{{\sl diff}}}\:{\rm e}^{-S}
    ~ \longrightarrow ~
   Z^{\rm discr}_{\rm pure}=\sum_{\rm T} {\rm e}^{-S}\;,
\label{disc}
\end{equation}  
with $S=\zeta\chi+\mu A$ in both cases if we identify $\chi$ with the
continuous and discrete version of the Euler characteristic and $A$ with
the surface area and the number of triangles, respectively.

A second assumption enters, when one replaces the summation over 
abstract triangulations by a summation over Feynamn graphs. By a duality
transformation each abstract triangulation can be identified with a
Feynman graph of an $N\times N$ hermitean matrix model with
cubic potential \cite{mm}. Not all Feynman graphs, however, correspond
to regular triangulations. The universality of the corresponding two
statistical ensembles is the subject of this letter.
In a continuum limit, where $A\!\rightarrow\!\infty$,
those graphs which from now on we refer to as {\it irregular}, even
dominate over the regular ones (see below, eq.\,(\ref{eq23})). 

The concrete case of 
pure ($c\!=\!0$) quantum gravity corresponds to a 1-matrix model and
the partition function for connected triangulations is given by the free
energy as the generating functional for connected vacuum graphs: 
\begin{equation}
Z^{\rm discr}_{\rm pure}  
       ~\sim~  
       F^{\rm matrix}_N(g) \equiv\frac{1}{N^2}\log Z^{\rm matrix}_N(g)  
   =\sum_{h,A}{\cal P}_h(A)\,g^AN^{-2h}\;,
\label{mat}
\end{equation}
where 
\begin{equation}
\label{eq3}
Z^{\rm matrix}_N(g)=\int {\rm d}^{N^2}\!\Phi\,\exp \left( -\frac{1}{2}
    {\rm tr}\, {\Phi}^2+
    \frac{g}{\sqrt{N}}{\rm tr}\, {\Phi}^3 \right) \;.
\end{equation}
Here $A$ denotes the number of vertices of the graph and ${\cal P}_h(A)$
is the number of graphs with given $A$ and genus $h$. As usual, 
(\ref{eq3}) is to be understood as a formal representation of an
asymptotic expansion in powers of $g$. Correspondingly, operations on
such expressions (taking the logarithm, differentiation, 
integration, etc.) are operations on formal power series expansions.
As (\ref{mat}) is a topological expansion in $1 \over N$ the
limit $N\rightarrow\infty$ results in the {\it planar} model to which we
want to restrict in the following.

The purpose of this letter is to show that universality of the
planar cubic model can be proven without knowing details of the 
model, as e.g.\ the spectral distribution of the matrix eigenvalues 
in the limit $N\!\rightarrow\!\infty$, which was required 
for the results
in \cite{bipz}, or the combinatorics of triangulations,
as it was used in \cite{Tutte}. The idea is to introduce new
couplings in the matrix model which can be adjusted using constraint
equations such that the irregular contributions in the perturbation 
expansion cancel. Without explicitly solving these equations they can
be used to relate the generating functional for regular graphs, 
$F^{\rm reg}(g)$, to the generating functional for all graphs, 
$F^{\rm all}(g)=F^{\rm matrix}_\infty(g)$, thereby proving
universality. Furthermore, given the behaviour of $F^{\rm all}(g)$
close to its singularity, our method allows to determine
the radius of convergence (i.e.\ the critical coupling) 
for $F^{\rm reg}(g)$.
  
We demonstrate this
method for the case $c\!=\!0$, where the logic of the procedure can be
illustrated most clearly. The extension to models with $0\!<\!c\!<\!1$ 
will in general be straightforward, and indeed has partly been used to
prove universality for the case $c\!=\!\frac{1}{2}$ \cite{burda}. 

In Sect.\ 2 we classify those irregular Feynman graphs which do not 
correspond to triangulations. In Sect.\ 3 we relate the generating
functional for regular graphs with the free
energy of a modified matrix model with renormalized couplings
by formulating constraint equations for these couplings. The proof
of universality follows in Sect.\ 4. In Sect.\ 5 we make use of some
known facts about the unregularized model to determine the radius
of convergence, i.e.\ the critical point, for the regularized model.

\section{Irregular Graphs}
The logic of our method is the construction of a
{\it regularized} model by elimination of all irregular graphs and 
the subsequent direct comparison of its critical behaviour with that 
of the original one. The first step thus 
consists in identifying the irregular graphs.

Consider graphs containing 1-point and non-trivial 2-point subgraphs: 
In the dual picture these correspond to situations 
where either two vertices of the same triangle are identified 
or two vertices of two different triangles are identified 
without identification of the connecting edges (links) -- see Fig.\ 1 
and 2. Those configurations are forbidden in the 
context of triangulations as discrete 2-dim.\ 
manifolds, see e.g.\,\cite{tri}. In turn, these are 
also the only irregularities that can arise from the planar cubic model 
considered here, i.e.\ we have a one-to-one correspondence between 
irregular graphs and graphs containing tadpoles 
and/or non-trivial 2-point subgraphs.

Note that this argumentation is independent of the value of $c$. It
depends, however, on the fact hat we are restricting to planar graphs
(there exist non-planar graphs without tadpoles and non-trivial 2-point
subgraphs which do not correspond to regular triangulations) and to 
graphs of valence 3.

\begin{figure}[htb]
\hspace*{3.0cm}
\psfig{figure=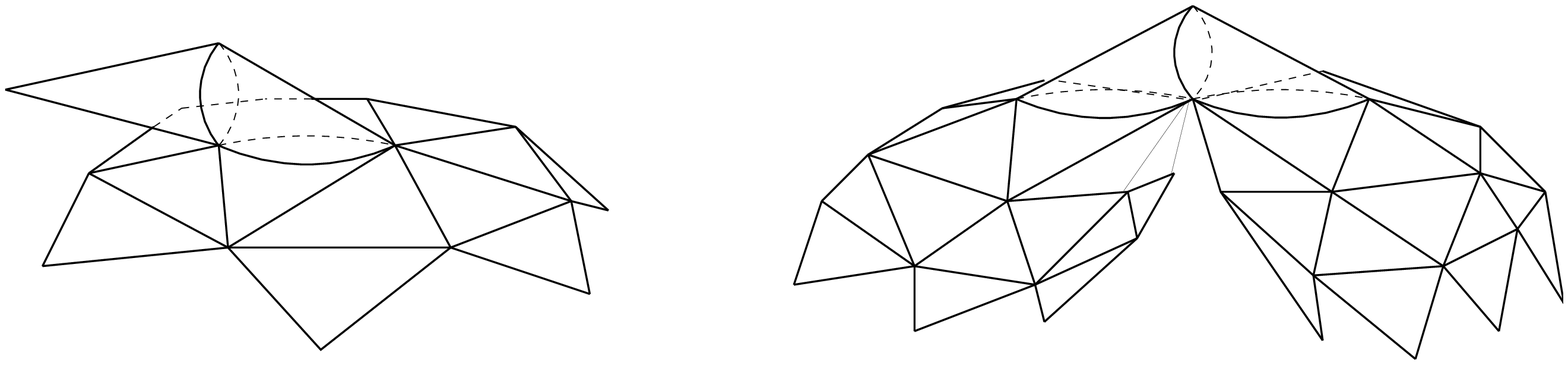,height=2.5cm,width=6.0cm}
{\caption{\sl Dual constructions of tadpole graphs}
\label{irreg2}}
\end{figure}

\begin{figure}[htb]
\hspace*{4.0cm}
\psfig{figure=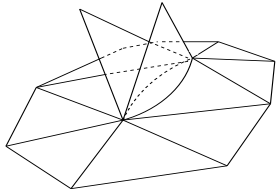,height=2.5cm,width=3.6cm}
{\caption{\sl General situation in the dual construction of a 
          2-point subgraph}\label{irreg1}}
\end{figure}

\section{Construction of the Regularized Model}

\begin{figure}[htb]
\hspace{2cm}
\psfig{figure=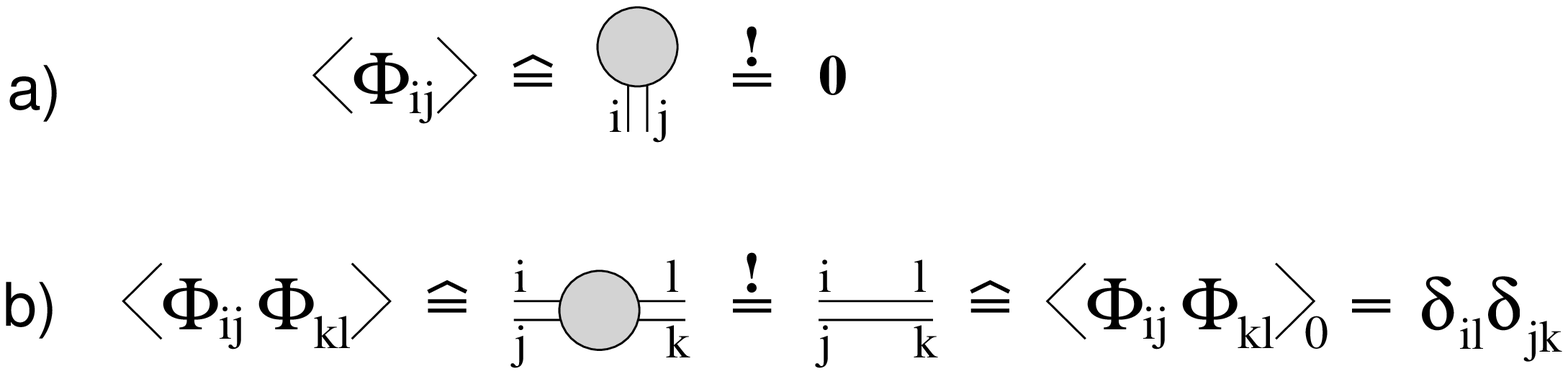,height=2.2cm,width=9.0cm}
{\caption{\sl Elimination of irregular graphs:
a) Tadpole elimination b) Elimination of self-energy contributions}
\label{rengra}}
\end{figure}

The construction of the regularized model, i.e.\ the generating
functional $F^{\rm reg}(g)$ for the numbers of regular triangulations,
can in general be achieved along the following steps:
\begin{enumerate}
\item 
Introduce a modified partition function (and corresponding free energy)
from a matrix action, which contains general couplings for those
contributions, on which one wants to put the contraints, i.e.\ 1-point-
and 2-point-functions:
\begin{eqnarray}
\nonumber
     S^{\rm mod}(\Phi) &=& -\frac{\alpha}{2}
      {\rm tr}\,{\Phi}^2
    +\frac{g}{\sqrt{N}} {\rm tr}\,{\Phi}^3+\rho\sqrt{N}{\rm tr}\,\Phi \\
    Z^{\rm mod}_N(g,\rho,\alpha) &=& 
    {\rm e}^{N^2 F^{\rm mod}(g,\rho,\alpha)} ~=~
       \int {\rm d}^{N^2}\!\Phi\: {\rm e}^{-S^{\rm mod}(\Phi)} \,,
\label{smod}
\end{eqnarray}
\item 
Impose two conditions on the free energy of the modified model
(\ref{smod}), where tadpoles are removed by setting the
1-point-function of the modified model equal to zero, i.e.
\begin{equation}
  \frac{\partial F^{\rm mod}(g,\rho,\alpha)}{\partial\rho}\,=\,0\;,
\label{rc1}
\end{equation}
and self-energy contributions represented by non-trivial 
2-point subgraphs are eliminated by 
assigning the value of the free propagator to the full 2-point function:
\begin{equation}
  \frac{\partial F^{\rm mod}(g,\rho,\alpha)}{\partial\alpha}\,=
  \,-\frac{1}{2}\,.
\label{rc2}
\end{equation}
A graphical representation of these two conditions is 
sketched in Fig.\ \ref{rengra}.

\item 
Evaluate the conditions (\ref{rc1}) and (\ref{rc2}) to find 
$\alpha(g)$ and $\rho(g)$. The free energy $F^{\rm reg}(g)$ is obtained
from $F^{\rm mod}(g,\rho,\alpha)$ 
by a Legendre transformation with respect to $\rho$ and $\alpha$,
\begin{eqnarray}
\label{leg}               
  F^{\rm reg}(g) &=& \left[ F^{\rm mod}(g,\rho,\alpha) -
       \frac{\partial F^{\rm mod}}{\partial \rho}\rho -  
       \frac{\partial F^{\rm mod}}{\partial \alpha}\alpha 
       \right]_{\rho=\rho(g),\alpha=\alpha(g)}\label{leg} \\
    & =& \left[ F^{\rm mod}(g,\rho,\alpha) +
            \frac{1}{2}\alpha 
       \right]_{\rho=\rho(g),\alpha=\alpha(g)} \;,
\nonumber
\end{eqnarray}
where we have made explicit use of the conditions (\ref{rc1}) and
(\ref{rc2}) in the second line. 
It will turn out that for the proof of universality it is not necessary
to know $F^{\rm mod}$ or $F^{\rm all}$ explicitly.
\end{enumerate}

\section{Proof of Universality}

For the proof of universality we have to show that $F^{\rm all}(g)$ and
$F^{\rm reg}(g)$ exhibit the same critical exponents when $g$
approaches its respective radius of convergence. Equivalently, we
will prove that the derivatives of both functions with respect to $g$
have the same critical behaviour.
First we note that
\[    \frac{{\rm d} F^{\rm reg}(g)}{{\rm d} g} ~=~ \left.
      \frac{\partial F^{\rm mod}(g,\rho,\alpha)}{\partial g}
      \right|_{\rho=\rho(g), \alpha=\alpha(g)} \;, \]
which is a general consequence of (\ref{leg}). On the other hand,
$F^{\rm mod}$ satisfies a generalized scaling relation which is easily
obtained from (\ref{smod}) by a change of variables $\Phi \rightarrow
\lambda \Phi$:
\[  F^{\rm mod}(g,\rho,\alpha) ~=~ \ln \lambda +
     F^{\rm mod}(\lambda^3 g, \lambda \rho, \lambda^2 \alpha)   \;. \]
Differentiating with respect to $\lambda$, setting $\lambda=1$, and
inserting the constraint equations (\ref{rc1}) and (\ref{rc2}) leads to
\begin{equation}
  \left.  \frac{\partial F^{\rm mod}(g,\alpha,\rho)}{\partial g}         
    \right|_{\rho=\rho(g), \alpha=\alpha(g)} ~=~
    \frac{\alpha(g) - 1 }{3 g}   \;. \label{hom}
\end{equation}
Thus, we have to show that $\alpha(g)$ has the same critical behaviour
as $\partial F^{\rm all}(g)/\partial g$. For this we will derive a
relation between $F^{\rm all}$ and $F^{\rm mod}$ and use the constraint
equations to obtain a relation between $\alpha$ and the derivative of
$F^{\rm all}$.

The partition function (\ref{smod}) is not changed by the introduction of
a new integration variable $\Phi=a\hat{\Phi}+b{\bf I}$ (where ${\bf I}$ 
denotes the $N\times N$ identity matrix). However,
the two parameters $a$ and $b$
may be tuned such that the action expressed in terms of $\hat{\Phi}$
contains no linear term, and the quadratic term appears with a factor
$-1/2$. In this way the modified model can be related to the original
matrix model and we obtain:
\begin{eqnarray}
\nonumber
        F^{\rm mod}(g,\rho,\alpha) &=& F^{\rm all}(k)\;-\;\frac{1}{4}\,
                \log\left(\alpha^2-12\,g\rho\right)\;+
       \;\frac{\alpha\!-\!\sqrt{\alpha^2\!-\!12g\rho}}{6\,g}
       \:\times 
       \\[0.3cm]
& & 
\times\:\left[\:\rho+
\frac{(\alpha\!-\!\sqrt{\alpha^2\!-\!12g\rho})^2}{36\,g}-
\frac{\alpha(\alpha\!-\!\sqrt{\alpha^2\!-12g\rho})}{12\,g}\:\right]\;,
\label{FF}
\end{eqnarray}
where
\begin{equation}
      k\equiv\,a^3g\,=\,\frac{g}{(\alpha^2-12g\rho)^{\frac{3}{4}}}\;.
\label{kg}
\end{equation}
  
Inserting this relation into the conditions (\ref{rc1}) and (\ref{rc2})
leads to 
\begin{eqnarray}
 \displaystyle{\frac{3\,g}{\alpha^2-12g\rho}\left(1+3k
         \frac{\partial F^{\rm all}(k)}
         {\partial k}\right)+\frac{\alpha-\sqrt{\alpha^2-12g\rho}}{6\,g}}
            &=& 0
\label{rb1}\\
\displaystyle{\frac{\alpha}{\alpha^2-12g\rho}\left(1+3k\frac{\partial
  F^{\rm all}(k)}{\partial k}\right)+\left(\frac{\alpha-\sqrt{\alpha^2-12g
            \rho}}{6\,g}\right)^2} &=& 1\;.
\label{rb2}
\end{eqnarray}
From these two conditions we can eliminate 
$\partial F^{\rm all}(k)/\partial k$ to obtain the first solution
\begin{equation}
         \rho(g)\;=\;-3\,g\;.
\label{rho}
\end{equation}
Without the explicit form of $F^{\rm all}(k)$ the
remaining condition cannot be solved to obtain 
$\alpha(g)$. However, for the proof of universality we only need to
confirm, that $\alpha(g)$ has the same critical behaviour for
$g\rightarrow g_{\rm c}$ as $\partial F^{\rm all}(k)/\partial k$ for
$k \rightarrow k_{\rm c}$. This follows immediately by inserting the
solution $\rho=-3g$ into the remaining constraint equation, 
say (\ref{rb1}), and making an expansion in $\delta k\equiv k-k_{\rm c}$.
For the equation to hold, the leading non-integer power of 
$\delta g=g-g_{\rm c}$ in $\alpha(g)$ has to be equal to the 
leading non-integer power of $\delta k$ 
in $\partial F^{\rm all}/\partial k$. This completes the proof of
universality.  

The whole procedure -- and thus the proof of universality -- immediately  
carries over to planar $c\!=\!0$ one-matrix-models with arbitrary even
potential of the order $2p$. In these models tadpoles are absent and the  
remaining renormalization of the 2-point function leads to an expression  
analogous to eq.\,(\ref{hom}), with the $3$ in the denominator replaced 
by $2p$.

Furthermore, note that relations (\ref{leg}), (\ref{FF}) and (\ref{rb1})
are independent of $N$. Furthermore, in the case of complex instead of
hermitean matrices, condition (\ref{rc2}) guarantees the
elimination of non-trivial 2-point subgraphs for {\it arbitrary 
topologies}. In general however, there will 
exist other irregularities not stemming from non-trivial
2-point subgraphs whose systematic elimination fails because their
classification is unclear.

\section{Critical Behaviour}

We now want to calculate the radius of convergence, $g_{\rm c}$, of the 
generating function of regular graphs, $F^{\rm reg}(g)$. This corresponds
to the critical point of the regularized model. For this we have to
know the radius of convergence, $k_{\rm c}$, of $F^{\rm all}(k)$ as well
as the leading coefficient in an expansion of $F^{\rm all}(k)$ 
around this critical point. We take these values from \cite{bipz},
for details of the calculation see also \cite{antje}:
\begin{eqnarray}
\label{kc}
       k_{\rm c}  &=&   \sqrt{\frac{1}{108\sqrt{3}}}  \\
\label{a1}       
  a_1  & \equiv & \left. \frac{\partial F^{\rm all}(k)}{\partial k} 
                   \right|_{k=k_{\rm c}} =   
       -\frac{2}{3\,k_{\rm c}}\left(\,5-3\sqrt{3\,}\,\right)\;.
\end{eqnarray} 
We now evaluate eq.\ (\ref{rb1}) at the critical point. To lowest order
we find
\begin{equation}
   0\,=\,\frac{1-3a_1k_{\rm c}}{\alpha_{\rm c}^2\!+36g_{\rm c}^2}
         +\frac{\alpha_{\rm c}\!-\!
       \sqrt{\alpha_{\rm c}^2\!+36g_{\rm c}^2}}{18\,g_{\rm c}^2}\;.
\end{equation}
With the given values for $k_{\rm c}$ (\ref{kc}) and $a_1$ (\ref{a1})
we therefore obtain an equation between $g_{\rm c}$ and 
$\alpha_{\rm c}=\alpha(g_{\rm c})$. A second independent relation between 
these two quantities is provided by (\ref{kg}), evaluated at the
critical point with the known solution $\rho=-3g$. From these two
equations we immediately obtain the new critical point to be
\begin{equation}
      g_{\rm c}\,=\,\sqrt{\frac{3}{256}}\;.
\label{gc}
\end{equation}
This value agrees with the result obtained by \cite{bipz} and 
\cite{Tutte}.

Let us finally add a comment on the fact, that universality is by
no means trivial. From the known cricital values we can deduce that
the number of graphs $n(A)$ as a function of the number of vertices
$A$ asymptotically grows like
\begin{equation}
   n(A) \stackrel{A\rightarrow\infty}{\longrightarrow}\:
   \sim\left(\frac{1}{k_{\rm c}}\right)^AA^{\kappa}\;,
\end{equation}
where $\kappa$ is the critical exponent proven to be universal.
$k_{\rm c}$ is the radius of convergence of the corresponding generating 
functional ($F^{\rm all}$ or $F^{\rm reg}$). (We should note that for
our choice of the matrix action (\ref{eq3}) the combinatorics of the
perturbation expansion leads to a factor of 3 for each vertex, i.e.\ after
a duality transformation one obtains an extra factor of 3 for each 
triangle in a 
triangulation. Our notation agrees with the one used by \cite{bipz} and
differs from \cite{Tutte} by this factor of 3 for each triangle.) 

Therefore, the ratio of the number of regular graphs 
$n^{\rm reg}(A)$ to the number of all graphs $n^{\rm all}(A)$ for
large values of $A$ is given by 
\begin{equation}
\label{eq23}
   \frac{n^{\rm reg}(A)}{n^{\rm all}(A)}
   \stackrel{A\rightarrow\infty}{\longrightarrow}\:\sim
   \left(\frac{256}{3\cdot108\sqrt{3}}\right)^{A/2}
   \stackrel{A\rightarrow\infty}{\longrightarrow}\:0\;.
\end{equation}
Thus, in the critical region the regular graphs considered as a subset
of all graphs represent a partition
of measure zero. So one cannot argue that
universality holds because the regular graphs ``dominate'' the ensemble.

\section{Summary and Outlook}

We presented a new and straightforward method of proving universality of 
the planar, cubic ($c\!=\!0$) matrix model with respect to the
elimination of graphs not corresponding to regular triangulations, 
thereby reproducing results of Br\'ezin et al.\ with, however, 
much less information needed about the original model. 
Our method also allows the determination of the new critical point 
from the knowledge of the old one.

An interesting generalization would be to models for which $c\ge 0$.
The extension of our method to models with $0\le c\le 1$ is in principle
possible and has partly been used in \cite{burda} for the case 
$c=1/2$. The situation for $c\geq 1$, however, is still unclear (see
also the references in \cite{Lattice}).


\newpage

\begin{thebibliography}{99}
\bibitem{Tutte} W.T.\,Tutte, Can.\,J.\,Math.\,14 (1962) 21.
\bibitem{bipz} E.\,Br\'ezin, C.\,Itzykson, G.\,Parisi, J.\,B.\,Zuber,
               Commun.\,Math.\,Phys.\,59 (1978) 35.
\bibitem{rev} D.\,J.\,Gross, T.\,Piran, S.\,Weinberg (eds.)\,,
               {\it Two dimensional Quantum Gravity and Random Surfaces},
               Jerusalem Winter School for Theoretical Physics, Vol.\,8
               (World Scientific,Singapore, 1991)\\
              F.\,David, {\it Simplicial Quantum Gravity and Random Lattices};
               SACLAY-preprint, T93/028.\\
              J.\,Ambj\o rn, {\it Quantization of Geometry}; NBI-HE-94-53.\\
              P.\,DiFrancesco, P.\,Ginsparg, J.\,Zinn-Justin, {\it 2d Gravity
               and Random Matrices}, Phys.Rep.\,254 (1995) 1
\bibitem{cont} J.\,L.\,Gervais, A.\,Neveu, Nucl.\,Phys.\,B\ 199 (1982) 59\\
               V.\,G.\,Knizhnik, A.\,M.\,Polyakov, A.\,B.\,Zamolodchikov,
               Mod.\,Phys.\,Lett.\,A\ 3 (1988) 819\\
               F.\,David, Mod.\,Phys.\,Lett.\,A\ 3 (1988) 1651\\
               J.\,Distler, H.\,Kawai, Nucl.\,Phys.\,B\ 329 (1989) 509
\bibitem{mm} G.'t Hooft, Nucl.\,Phys.\,B 72 (1974) 461\\
             F.\,David, Nucl.\,Phys.\,B 257 (1985) 45 \\
             V.\,Kazakov, Phys.\,Lett.\,B 150 (1985) 282 \\
             J.\,Ambj\o rn, B.\,Durhuus, J.\,Fr\"ohlich, Nucl.\,Phys.\,B 259 
             (1985) 433       
\bibitem{burda} Z.\,Burda, J.\,Jurkiewicz, Act.\,Phys.\,Pol.\,B 20 (1989)  
949               
\bibitem{tri} Encyclopedic Dictionary of Mathematics, MIT Press, 1987.
\bibitem{antje} A.\,Schneider, diploma thesis, 
Universit\"at des Saarlandes (1996)
\bibitem{Lattice} M.\,Bowick; {\it Random Surfaces and Lattice Gravity},
   in LATTICE\,97;
   Nucl.\,Phys.\,B (Proc.\ Suppl.) 63 A-C (1998) 77.  
\end{thebibliography}
\end{document}